\documentclass[a4paper]{article}
\usepackage{graphicx}
\usepackage{amsmath}
\usepackage{hyperref}
\usepackage{color}

\begin{document}

\begin{center}

{\bf \Large Opinion polarization in the Receipt-Accept-Sample model}\\[5mm]

{\large  Krzysztof Ku{\l}akowski }\\[3mm]

{\em

Faculty of Physics and Applied Computer Science,

AGH University of Science and Technology,

al. Mickiewicza 30, PL-30059 Krak\'ow, Poland

}

\bigskip

$^*${\tt kulakowski@novell.ftj.agh.edu.pl}

\bigskip

\today

\end{center}

\begin{abstract}
The Zaller theory of opinion formation is reformulated with one free parameter $\mu$, which measures the largest
possible ideological distance which can be made by a citizen in one mental step. Our numerical results show the 
transient effects: {\it i)} the political awareness, measured by the number of 
received messages, increases with time first exponentially, later linearly; {\it ii)} for small $\mu$ 
correlations are present between previously and newly received messages; {\it iii)} these correlation lead to 
a hyperdiffusion effect in the space of attitudes of messages. Citizens with small $\mu$ are more prone to extremal 
opinions.

\end{abstract}

%\noindent

%{\em PACS numbers:} 89.65.-s, 64.90.+b

%\noindent

%{\em Keywords:} random networks; phase transitions; 

%% #####################################################################

\section{Introduction}

%% #####################################################################

It is difficult to overestimate the importance of the public opinion (PO) in our life \cite{lipp}. In all 
political systems, PO transfers the traditional receipts of solving problems to subsequent 
generations. Its overwhelming influence is expressed by the democratic laws, which transfer the ultimate 
decisions to the voters. In modern societies, the role of PO is amplified by mass media
- "the fourth power", which is sometimes more authoritary that each of the remaining three. (Recently
however, the list of agents at the market of information is enlarged due to the Internet, and the monopoly 
of the mass media is somewhat weakened \cite{cast}.) The very importance of PO was not overlooked by the 
sociophysicists, and several models have been created to capture its dynamics - for recent reviews 
see \cite{stau,cafl}. On the other hand, the subject has been modeled mathematically also by social 
scientists, and their models are of particular interest for the sociophysicists, as those models can 
be expected to have a firm sociological basis. \\

The subject of this text is the model of the opinion dynamics, formulated by the political scientist 
John Zaller and published for the first time in 1992 \cite{zall}. The book is cited about 1000 times 
\cite{gosc}, but only one citation \cite{bocc} can be found in the arXiv database \cite{arxi}, where 
sociophysicists met. The book is devoted to the explanation and the verification of the model, which 
is clearly exposed on its first page \cite{zall}:\\

"The ideas necessary to accomplish this integration are few and surprisingly simple. 
The first is that citizens vary in their habitual attention to politics and hence 
in their exposure to political information and argumentation in the media. The second 
is that people are able to react critically to the arguments they encounter only to 
the extent that they are knowledgeable about political affairs. The third is that 
citizens do not typically carry around in their heads fixed attitudes on every issue 
on which a pollster may happen to inquire; rather, they construct "opinion statements" 
on the fly as they confront each new issue. The fourth is that, in constructing their
opinion statements, people make greatest use of ideas that are, for one reason or 
another, most immediately salient to them..."\\

Zaller termed his model 'Receipt-Accept-Sample', or RAS, and we shall use this abbreviation below in the text. 
Also we accept his term 'citizen' for a social actor. The original formulation contains at least 8 free parameters. 
This is convenient when we struggle for accordance with experimental data, but less handy for 
somebody interested in the general behaviour of the model. Our method here is to simplify
the mathematical formulation of the RAS model as much as possible, preserving the sociological content of the  
four postulates cited above. The aim of this paper is to investigate the distribution of opinions about a model 
issue. Our approach to the Zaller model profits much from the continuous opinion dynamics, as described
by Deffuant et al. in 2000 \cite{defw}. There, the core idea is that other opinions are taken into account by an agent if the
distance from their content to his actual attitude is not larger than some threshold value. Here, the 
distance $d$ is measured from a new message to the closest point on the plane of opinions, occupied by the agent in the past.
This distance is compared with the threshold parameter $\mu$; if $d<\mu$ the message is received, but it is ignored if $d>\mu$.  \\

The Zaller's mathematical formulation of the RAS model is briefly reviewed in our Section 2. In Section 3 we describe 
its simplified version proposed in our model. Last two sections are devoted to the numerical results and their discussion. 

\section{The original mathematical formulation}

The postulates listed above are expressed as formal axioms which constitute the RAS model \cite{zall}. On this 
basis, a mathematical construction is built as follows:\\

1. Each citizen $i$ is endowed with the political awareness of a given value $W_i$.\\
2. The probability of receipt a message relevant for the opinion formation increases
with his awareness according to a sigmoidal function $f$

\begin{equation}
f(W_i;a_0,a_1)=1-\frac{1}{1+f+\exp(a_0+a_1W_i)}
\end{equation}
where $f$ - floor parameter, which marks a minimum level of reception, $a_0$ - a coefficient 
which designates the intensity of a message, and $a_1$ - a coefficient which designates 
strength of a relationship between awareness and reception. \\

3. Provided that a message is received by $i$, the probability of its acceptance decreases 
with the awareness $W_i$ and it is given by another sigmoidal function $g$

\begin{equation}
g(W_i,P_i;b_0,b_1,b_2)=\frac{1}{1+\exp(-b_0-b_1W_i-b_2P_i)}
\end{equation}
where $b_0$ - a coefficient which designates the difficulty or credibility of the message,
$b_1$ - a coefficient which designates the effect of awareness on resistance to persuasion, 
$b_2$ is a coefficient which designates the effect of the predisposition on resistance to 
persuasion, and $P_i$ measures the predisposition of $i$ to accept the message, and it depends
on the ideological relation of $i$ to the message content.\\

4. The probability of acceptance of the message and a henceforth change of attitude is equal 
to the product $Pr= f(W_i;a_0,a_1)g(W_i,P_i;b_0,b_1,b_2)$. \\

5. The latter expression can be used to construct a kind of one-way Master equation for the 
probability $Prob$ of a given opinion $Opi$, provided that at $t=0$, the baseline opinion was $Bas$

\begin{equation}
Prob(Opinion)_t=Prob(Bas)+(1-Prob(Bas))*Pr*Dum_t
\end{equation}
where $Dum_t=0$ and 1 at initial and final time, respectively. The final opinion is expected to 
be expressed in an opinion statement in a poll.\\

6. Additional consideration is due to a situation when the citizen is a subject of two streams of 
opposite messages. In this case, the decision on the content the opinion statement is undertaken 
according to opinions most immediately accessible in a citizen's mind. The probability to recall
a previously accepted opinion is given by yet another sigmoidal function $h$

\begin{equation}
h(W_i;c_0,c_1)=1-\frac{1}{1+\exp(c_0+c_1W_i)}
\end{equation}
where the parameters $c_0$ and $c_1$ are analogous to $a_0$ and $a_1$. \\

7. To be specific, let us consider two opposite attitudes, say pro- and antiwar (P or A) ones
\cite{zall}. Let us also assume that our statistical citizen expresses some opinion statement 
on this issue; actually, a separate expression is given in \cite{zall} for the probability of this
fact. The probability of the prowar opinion statement $p_p$ expressed by citizen $i$ is supposed to be

\begin{equation}
p_p=\frac{S_p(i)}{S_a(i)+S_p(i)}
\end{equation}
where $S_p$ ($S_a$) is the accessibility of the prowar (antiwar) messages in the citizen's mind.
These accessibilities are given as

\begin{equation}
S_p(i)=\sum_t f(W_i;a_0,a_1)g(W_i,P_i;b_0,b_1,b_2)h(W_i;c_0,c_1)
\end{equation}
where $t$ is time and the sum is performed over a given time period (say, two recent years).
In this sum, the constants $a_0$, $b_1$ and $b_2$ depend on time $t$, and the constants $a_0$ and 
$b_0$ depend on the message (prowar or antiwar).\\

8. Some additional assumptions (all the messages of the same intensity) allow to find all the 
coefficients from the fitting of the theoretical curves to the poll data.\\

\section{Our formulation}

To investigate, as we intend here, the distribution of opinions, it is necessary to postulate 
how the opinions vary in time. On the contrary to most sociophysical approaches, the RAS model \cite{zall} does not
take into account direct interactions between citizens; it is only the influence of media what is taken
into account. This does not preclude the possibility that some citizens play the role of media, which could
be taken into account in future research. Here we are going to preserve the one-particle character of the
Zaller model. Our interest is in the time dynamics of PO with respect to a given issue.\\

In particular we are interested in a possible sequence of events when a newborn citizen starts to hear
political news. His initial awareness is close to zero, but he is indiscriminative in his attitude and he 
accepts any news as typical and normal. In this sense, his initial acceptance is large. Being a subject of a random 
stream of messages, initially he is 
not able to receive most of them, because they seem to him to be too sophisticated. There are some, however, apparently
addressed to citizens like him: full of emotions, which clearly divide the world into good and bad, expressed by 
somebody authoritative but young. Our citizen captures such a message and learns to distinguish it from other
messages. His political education just started, and his awareness increases a little bit. Simultaneously, his 
acceptance is strongly reduced. At least in the near future he will be willing to identify his opinion with this 'his first' 
message. In physics we like to term such events as 'spontaneous symmetry breaking' \cite{ball}.\\

To capture the evolution of the citizen's understanding, we need {\it i)} a history-dependent awareness, 
{\it ii)} a time series of messages which vary in their content with respect to some set of issues which we consider 
to be salient in a given society. 
In fact, we do not need anything more to indicate, that there is a positive feedback between the political
attitude and the character of newly received informations. On the contrary to Ref. \cite{bocc}, we do not need a 
Gaussian or any other distribution of awareness, as this distribution should be a result of the calculation.
Also, we consciously do not contribute to the discussion if a given political orientation is correlated 
with the awareness \cite{bio}. Instead, we will show that a citizen can be randomly trapped by a series of 
messages close to each other till the time when his attitude is firmly established.\\

The mathematical formulation of our version of the RAS model is then as follows:\\
\noindent

1. At the beginning, the political awareness of each citizen is equal to one. Later it is represented by the number
$n(t)$ of different opinions/messages received till time $t$. Each citizen starts with one received message, 
placed at the centre of coordination.\\

2. Each new message is represented as a point on a plane, with coordinates $(x,y)$. The plane plays the role of the 
space of main attitudes with respect to some salient issues, for example 'safety vs freedom' and 'free trade vs 
welfare state'. Simultaneously, each message is represented by a point at this plane. The actual value of the 
dimensionality of this space is of minor importance, except the condition that it is larger than one. When the 
dimensionality $D$ is two or larger, the size of the $(D-1)$-dimensional circumference of the occupied area increases 
with its size. This means, that the ability to receive new messages increases with the awareness. \\

3. The ability to receive a new message is a decreasing function $f$ of the distance $d$ between the point $(x,y)$
and the closest point received in the past. For simplicity we adopt the threshold function $f=\Theta(\mu-d)$ \cite{defw}.
Here $\mu$ is a parameter, which can be roughly interpreted as the capacity to receive new messages. Receipt of a new message 
is equivalent to a visit at the point $(x,y)$ assigned to this message. The starting point is placed at the 
coordination center.\\

4. The spatial distribution of received messages can be used to calculate the probabilities of the opinion statements in 
a similar way, as Zaller did in Eq. (5). In our case, the probabilities are calculated from the political contents 
of the previously received messages, i.e. the position of the points which represents
the messages. For an exemplary issue defined by the $y$ axis as the boundary between the opinions 'pro' and 'anti', 
the message weight is just its $x$ coordinate. Having chosen $x>0$ as 'pro' and $x<0$ as 'contra' we can calculate the 
respective probability 'pro' as $p$

\begin{equation}
p=\frac{\sum_{x>0}x(t)}{\sum \mid x(t)\mid}
\end{equation}
\\

5. Other issues can be visualised as other axis, not necessarily orthogonal to the plane of salient issues, defined above. 
Opinions on those other issues can be formed on the basis of the projection of the new axis to the 'salient' plane.\\

6. Once a citizen has a given attitude 'pro' along the OX axis, i.e. $p>0.5$, this attitude can be neutralized by 
receipt a given number of messages with $x<0$. It is natural to set $p=0$ as a neutral attitude. Then, the number 
of messages to neutralize $p$ can be roughly evaluated as $(2p-1)n(t)$.\\

It seems to this author, that this formulation fulfils the content of 
the postulates of Zaller's theory. As mentioned above, the distribution of the awareness
appears as a natural consequence of the procedure listed above. As we are going to demonstrate,
the area around received messages varies from one citizen to another. Simultaneously, the circumference 
of this area is a measure of the amount of ideas which can be accessible by the citizen in a near future.
A good school is where young minds are gradually fed by new ideas, without prejudices towards 
this or other direction. On the contrary, if a citizen is indocrinated by only one idea, 
he is not able to receive anything else; in our model, such brainwashing is equivalent to the 
case when subsequent messages either are close to the current position, or too far from it to
be received. Actually, the stream of messages we encounter in our life is not completely random, 
but it depends on our intellectual environment. At the moment, however, we do not construct a 
theory of the whole society, then our analysis is limited to citizens and does not embrace 
the media. 

We hope that the arguments given in the preceding paragraph allow to state that the first and the 
second postulate of Zaller's theory: that the citizens vary in their awareness, and that 
they are able to take into account only those ideas which they are knowledgeable about. The essence 
of the third and fourth postulates is captured in the point that the opinion statement is formulated 
on the basis on the recent trajectory, but separately for each new issue formulated in a poll.
Here we do not concentrate much on the sampling process, when a new issue is formulated. According to Zaller, opinions
on such new issues are formulated on the basis of other, most salient issues. To refer to an opinion 
on a new issue, a relation should be determined between this new issue and the old (salient) one. 
We imagine that such a relation can be constructed in a three-dimensional space; if a new axis is orthogonal 
to the old ones, nothing can be stated on a new issue. However, if their angle is different from $\pi /2$,
information on the distribution of opinions can be drawn by means of a geometrical projection.

The selected form of the function $f$ is adopted from the Deffuant model. In general, any decreasing function 
$f(x)$ fulfils the condition that messages more distant are less likely to be received. On the other hand,
one of the postulates by Zaller indicates that some messages can be received but some others can not,
and that the ability to receive them increases with the awareness of the receiving citizen. In our formulation, 
the awareness increases with the area around previously received messages. When this area is larger, more new 
messages can be received.

\section{Numerical results}

In a generic case, the awareness of a citizen increases with time at first exponentially, later it becomes a linear 
function of time. An example of the time dependence of the awareness is shown in Fig. 1. At the later 
stage all incoming messages appear to be at the already known area.
Numerically, the distance between a new point and some already received points is shorter than the threshold $\mu$;
then, the new point is also received. This means that the number of received messages increases by one
at each time step. At the earlier stage, however, the time to receive a new message is longer. Namely,
the probability to receive the second message is equal to $s=\pi \mu^2/4$, where the factor 1/4 comes from the fact
that new messages appear as points placed randomly at the square $2\times 2$. Then the probability $P(n)$ 
to pass $n$ new mesages until the receipt of the $n+1$-th one and the increase of the awareness 
to $W=2$ is $P(n)=s(1-s)^n$. The time to receive the third message depends on where the first received point 
is placed, and should be averaged over its possible positions. Similarly the distributions of waiting time to get 
subsequent messages depend on the positions of the points which refer to previously received messages.
The numerical calculations indicate, that these functions also decrease exponentially with $n$, and the rate 
of decrease is larger for higher values of the awareness. The plots are shown in Fig. 2. This means in particular, 
that the mean time to receive a new message decreases with the awareness. 

\begin{figure}
\begin{center}
\includegraphics[scale=0.45,angle=-90]{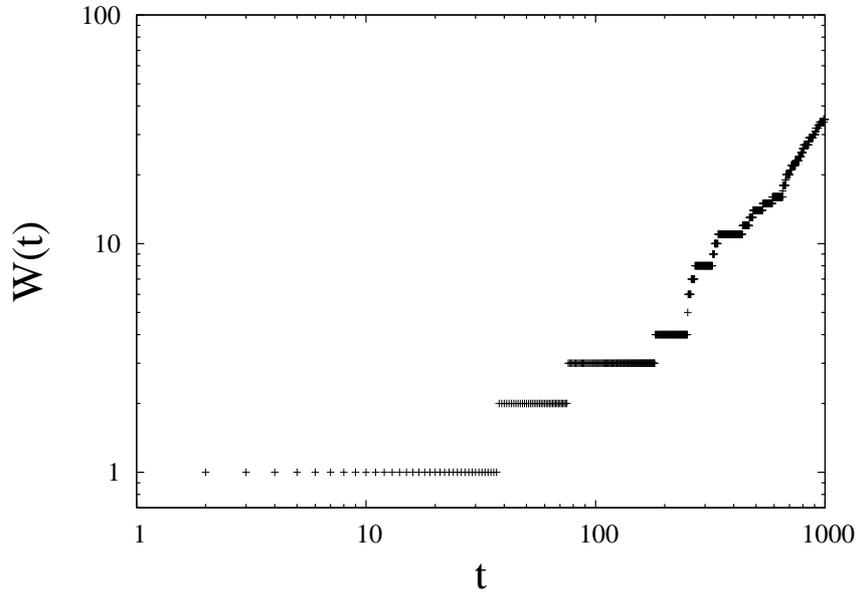}\\
\end{center}
\caption{Example for the time dependence of the awareness $W(t)$.
}
\end{figure}

\begin{figure}
\begin{center}
\includegraphics[scale=0.45,angle=-90]{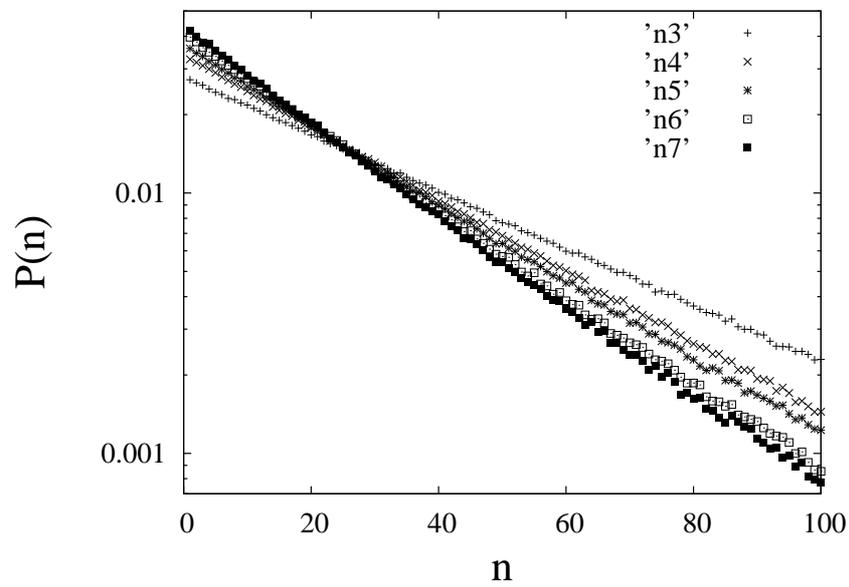}\\
\end{center}
\caption{The probability distributions $P(n)$ to wait $n$ timesteps until an increase of the awareness $W$ to the values 3, 4, 5, 6 and 7.
}
\end{figure}

Being (socio)physicists, we are tempted to investigate the geometrical aspect of the correlation. For this purpose 
we calculate the time dependence of the square $r^2$ of the sum of coordinates of the messages 
received till a given time against time. Note that in this way a point is taken into account several times, 
if no messages were received after the message which it represents. Subsequent points are not correlated in two 
cases: first, if the whole plane is already densely occupied by previously received messages, and second - for 
large $\mu$, where any new message is received, despite its position. In the latter case, the slope of $r^2$ against 
time should be 1 in the log-log scale in accordance with the diffusion law. Our numerical results 
indicate, that there are two transient times for small $\mu$: $t_1(\mu)$ and $t_2(\mu)$. For $t<t_1(\mu)$ the slope is 
larger than 1 and reaches 2.5. For $t_1(\mu)<t<t_2(\mu)$ the slope is smaller than 1. Finally, for $t>t_2(\mu)$ the slope is 
1 for each $\mu$. When $\mu$ increases, both $t_1(\mu)$ and $t_2(\mu)$ decrease and merge.  
In this way, we observe a kind of cross-overs in time from the hyperdiffusion to the subdiffusion and then to the normal 
diffusion. These results are presented in Fig. 3. The plots are averaged over 1000 trajectories. The results are the same 
if we limit the area where new points appear to a circle with radius equal to 1.0.

\begin{figure}
\begin{center}
\includegraphics[scale=0.45,angle=-90]{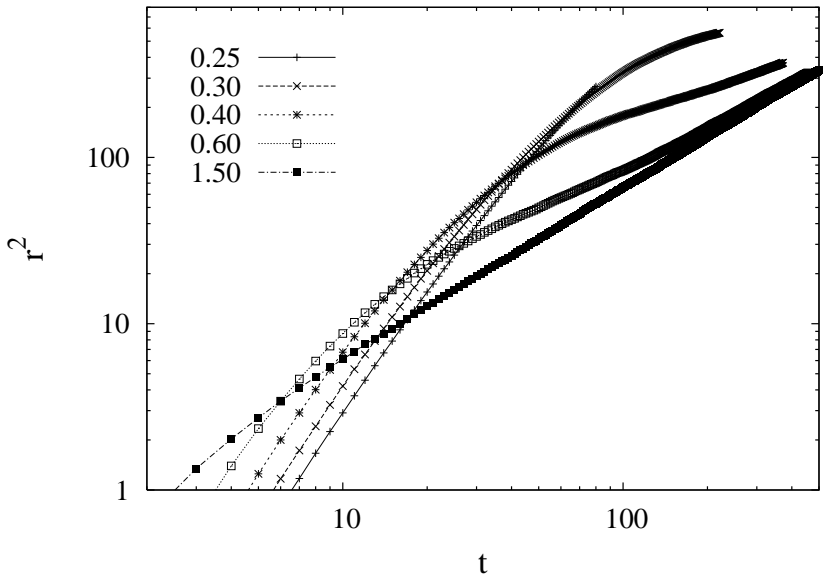}\\
\end{center}
\caption{Time dependence of the squared sum of positions of all received messages for $\mu$ = 0.25, 0.3, 0.4, 0.6 and 1.5. 
The slope of the curves except the one for $\mu$=1.5 is initially larger than 1.
}
\end{figure}

A characteristic feature of the model is that subsequently received messages are correlated in their ideological content. 
In our geometrical realisation this correlation is expressed as the correlation between positions of subsequently received messages.
Once a new message is selected, the area around received messages gets widened towards a given direction. This raises the possibility of 
correlations between subsequent messages. Then, we calculate this correlation for the direction $x$ - the results are shown in Fig. 4. We use the 
expression for non-stationary correlations, i.e. 

\begin{equation}
W_x=\frac{\langle x(t)x(t+1)\rangle-\langle x(t)\rangle \langle x(t+1)\rangle}{\sqrt{Var(x(t)) Var(x(t+1))}}
\end{equation}
\\

This correlation is larger for small values of $\mu$. When the whole area is covered, the correlation
disappears. It is obvious, that the correlation lifetimes increase when $\mu$ decreases. The results should not depend on the selected 
direction; the $OX$ axis is chosen for simplicity of the parametrization. The effect of correlations is analogous to the viscous fingering in the snowflake formation \cite{fract}, where randomly attached molecules increase the probability of attaching further molecules at the same area, and the curvature of the surface increases.

\begin{figure}
\begin{center}
\includegraphics[scale=0.45,angle=-90]{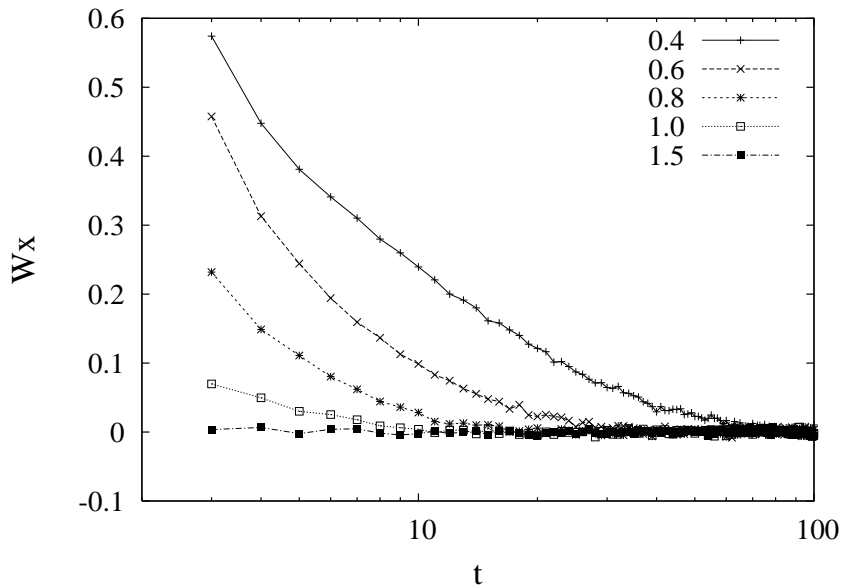}\\
\end{center}
\caption{Time dependence of the correlation $Wx$ between two subsequently received points (messages) for $\mu$ = 0.4, 0.6, 0.8, 1.0 and 1.5.
}
\end{figure}

In Fig. 5 we show the probability distribution $P(p)$ of these probabilities at a not-too-long time moment, i.e. after receipt of 500 messages.
As we see, the obtained plot is close to the Gaussian distribution for $\mu >0.4$, but its shape deviates from Gaussian for smaller $\mu$ because of the  limitation of the variable $p$ to the range (0,1). None of these plots shows a fat tail. Note that different
trajectories contribute to the average with their different numbers of received messages. We also checked that in time, the distribution gets narrow;
the time dependence of the variance of $P(p)$ decreases, as shown in Fig. 6. This narrowing can be interpreted as follows: as citizens get
more complete information, their opinions are clarified. The final mean probability $<p>=0.5$; this follows from the initial symmetry of the 
distribution, which is preserved by the dynamics. 

\begin{figure}
\begin{center}
\includegraphics[scale=0.45,angle=-90]{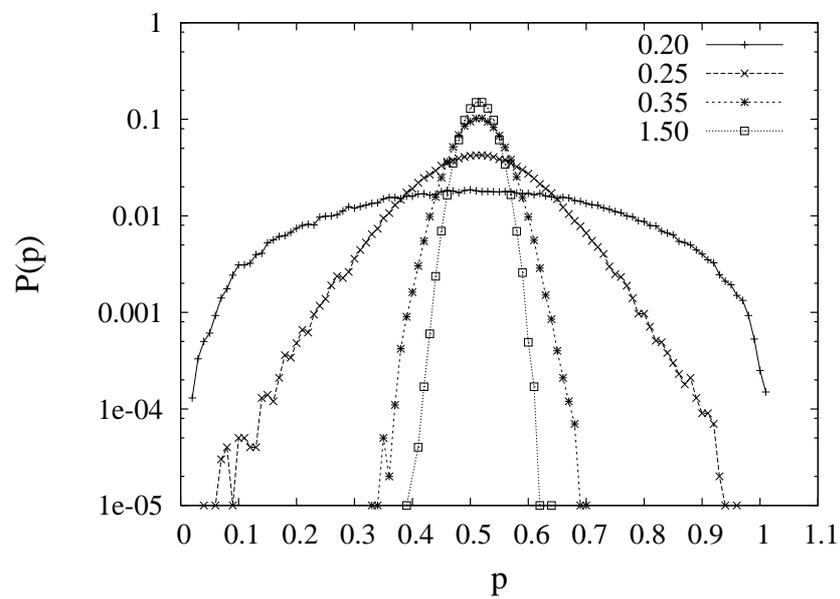}\\
\end{center}
\caption{The probability distribution $P(p)$ that the opinion statement of randomly selected agent is equivalent to accept a given issue with 
probability $p$. The curves are obtained after 500 messages are received, for $\mu$=0.2, 0.25, 0.35 and 1.5 from the widest to the most narrow curve,
respectively, and averaged over $10^5$ citizens.
}
\end{figure}

Let us remind that the factor $\mu$ describes the largest possible distance between messages, from the previously to the one newly received.
It is somewhat surprising that opinions of the citizens which understand less swiftly,. i.e. of those with smaller $\mu$, is temporarily more wide.
This effect is visible in Fig. 5, where the distribution of probabilities is more wide for smaller value of $\mu$. The reason is the correlation between subsequently received messages, which is larger for smaller $\mu$.

\section{Discussion}

The assumption most essential here is the geometrical-like distance between messages, determined by their political content. Next assumption
correlates the distance between a new message and previously received messages with the receipt of the new message. The latter seems natural,
the former can seem doubtful. However, separating out the geometrical considerations, we are left with the conjecture that messages
with the content far from anything previously received are received less likely. This conjecture seems to be quite natural. 

It is possible to continue this kind of reasoning. Our main goal is to demonstrate, that opinions of people endowed 
with smaller factor $\mu$ are more spread, i. e. there are more extremal opinions, than for the people with larger $\mu$. 
To falsify this theory one should demonstrate that the result should be the opposite, i.e. the capacity to receive new 
informations is positively correlated with the extremal opinions. It seems to this author that this is not the case. 
On the other hand, in our model new messages appear in a limited area. Then, the model is appropriate to a given area 
of knowledge (here the square $2\times 2$), which sooner or later will be covered by received messages. We think that 
capable citizens spread their fields of interest beyond the boundaries which limit thoughts of less endowed people. 
Then in new areas everybody is a freshman at least at the beginning of their exploration.

Our result that opinions of people with smaller $\mu$ are more spread is akin to the result that opinions of people 
with smaller threshold parameter are more spread (Figures 1 and 2 in Ref. \cite{defw}). Although the outcome of both 
results is the same, their origin is different. The Authors of Ref. \cite{defw} assumed the homogeneous initial distribution
of opinions. There, smaller threshold means smaller adaptation and smaller mobility in the opinion space. As a consequence,
the distribution remains wide. In our case, the initial distribution is concentrated at the centre. Smaller $\mu$ means
smaller adaptation and smaller mobility, but larger correlations. As a consequence, the distribution gets wider.

\begin{figure}
\begin{center}
\includegraphics[scale=0.45,angle=-90]{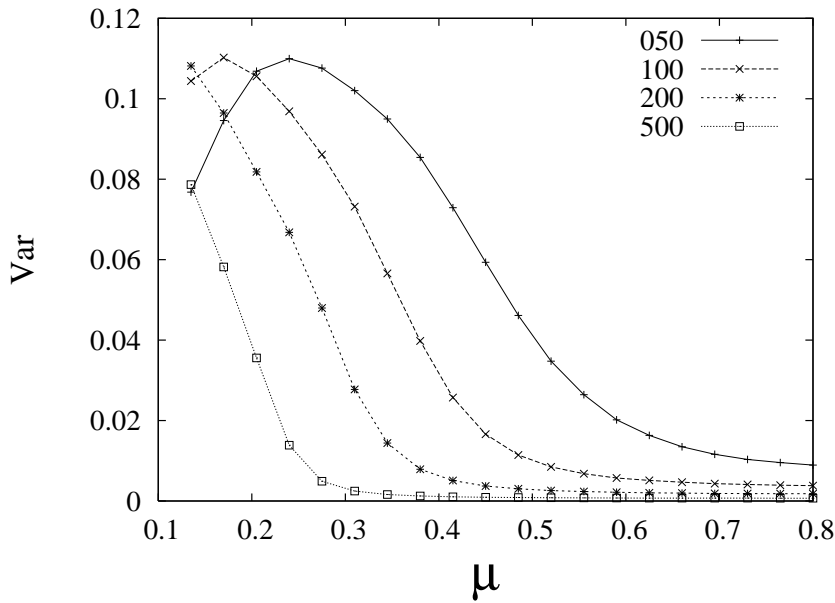}\\
\end{center}
\caption{The variance of the probability distribution $P(p)$ against the capacity parameter $\mu$, for various times of calculation: from 50 to 500 timesteps.
}
\end{figure}

We found that the awareness, measured by the number of received messages, increases with time exponentially, later linearly. The seemingly exponential part reveals a substructure, which can be of interest for statistically oriented minds, but its existence 
relies on the assumption that messages can be represented by discrete points. It is possible that their representation as, say, extended 
excitations in a network of ideas could be more fruitful. Other result are indications of a hyperdiffusion in the space of messages. The effect is transient and it disappears when the considered area becomes known. In this transient time and for small $\mu$, positive correlations are found between the content of subsequently received messages. The transient time decreases with $\mu$; more capable citizens are quicker.

One of the conclusions by Zaller was that the acceptance decreases with the awareness. In our formulation, the equivalent of non-acceptance is 
the number of messages necessary to neutralize a given attitude. For a given $p$, this number increases with the awareness $n$, as remarked 
in point 6 of our formulation. This point does not preclude the mean decrease of non-acceptance with time, as shown in Fig. 6. However, more 
detailed discussion of this question should take into account some initial distribution of attitudes. Here we concentrate on the influence
of $\mu$ on the opinion spread; then we purposefully neglect this initial distribution, placing initially each citizen at the coordination centre. 

We feel that the weak part of our formulation is that the messages previously received do not lead in the model to 
a coherent picture, but they stand by in minds till the end of the simulation. A partial remedy could be to add 
another parameter of forgetting old messages in time, as was done for example in the Bonabeau model \cite{bona}. 
Zaller proposed to take into account only some last part of the simulation. This effect is under investigation. 
Even more essential process 
disregarded here as in Ref. \cite{zall} is the contact between people. This mechanism of the opinion formation was 
discussed by many sociophysicists \cite{sznajd,krahe,weis,gala}; we recommend \cite{cafl} as the current review. 
The next challenge is to build the interpersonal mechanism into 
the RAS theory. For this purpose, the Deffuant model \cite{defw,weis} seems most natural. This model relies on 
the assumption that we react to opinions of other people if they are not too far from our initial beliefs. 
Our formulation of the RAS theory explores the same idea.

\bigskip

{\bf Acknowledgements.} I am grateful to two Anonymous Referees for their helpful remarks. The calculations were performed in the ACK Cyfronet, Cracow, grants No. MNiSW /SGI3700 /AGH /030/ 2007 and  MNiSW /IBM BC HS21 /AGH /030 /2007. The text is dedicated to my wife Magda.

\bigskip

%\section*{Acknowledgements}

\end{document}